\documentclass[journal]{article}
\usepackage{amsfonts,amssymb,anysize}
\usepackage{times,mathptm}
\usepackage{amsmath,amsthm}
\usepackage{algorithmic,algorithm}
\usepackage{graphicx}
\usepackage{hyperref}
\marginsize{0.9in}{0.9in}{0.6in}{0.8in}

\newtheorem{lemma}{Lemma}

\begin{document}

\title{A Study of Optimal 4-bit Reversible Toffoli Circuits and Their Synthesis
\thanks{This work has been submitted to the IEEE for
possible publication. Copyright may be transferred without notice, after
which this version may no longer be accessible.}}

\author{Oleg~Golubitsky and Dmitri~Maslov
\thanks{O. Golubitsky is with Google Inc., Waterloo, ON, Canada.}
\thanks{D. Maslov is with the Institute for Quantum Computing, University of Waterloo,
Waterloo, ON, N2L 3G1, Canada, email: {dmitri.maslov@gmail.com}.}
}


\maketitle

\begin{abstract}
Optimal synthesis of reversible functions is a non-trivial problem.  One of the major limiting
factors in computing such circuits is the sheer number of reversible functions.  Even restricting
synthesis to 4-bit reversible functions results in a huge search space (16! $\approx$ $2^{44}$
functions).  The output of such a search alone, counting only the space required to list Toffoli
gates for every function, would require over 100 terabytes of storage.

In this paper, we present two algorithms: one, that synthesizes an optimal
circuit for any 4-bit reversible specification, and another that synthesizes all optimal implementations.  
We employ several techniques to make the problem tractable. 
We report results from several experiments, including synthesis of all optimal 4-bit permutations, 
synthesis of random 4-bit permutations, optimal synthesis of all 4-bit linear reversible circuits,
synthesis of existing benchmark functions; we compose a list of the hardest permutations to synthesize, 
and show distribution of optimal circuits.  We further illustrate that our proposed approach may be extended 
to accommodate physical constraints via reporting LNN-optimal reversible circuits.  Our results
have important implications in the design and optimization of reversible and quantum circuits, testing circuit
synthesis heuristics, and performing experiments in the area of quantum information processing.
\end{abstract}



\section{Introduction}
To the best of our knowledge, at present, physically reversible technologies are found only in 
the quantum domain \cite{bk:nc}. However, ``quantum'' unites several technological approaches 
to information processing, including ion traps, optics, superconducting, spin-based and 
cavity-based technologies \cite{bk:nc}. Of those, trapped ions \cite{ar:s-k} and
liquid state NMR (Nuclear Magnetic Resonance) \cite{ar:n} are two of the most
developed quantum technologies targeted for computation in the circuit model (as opposed to communication
or adiabatic computing). These technologies allow computations over a set of 8 qubits and 
12 qubits, correspondingly.

Reversible circuits are an important class of computations that need to be performed efficiently for the 
purpose of efficient quantum computation. Multiple quantum algorithms contain arithmetic units such as 
adders, multipliers, exponentiation, comparators, quantum register shifts and permutations, that are best 
viewed as reversible circuits.  Moreover, reversible circuits are indispensable in quantum error 
correction \cite{bk:nc}. Often, the efficiency of the reversible implementation is the bottleneck of a 
quantum algorithm (e.g., integer factoring and discrete logarithm \cite{ar:s}) or even a class of quantum 
circuits (e.g., stabilizer circuits \cite{ar:ag}).

In this paper, we report algorithms that find optimal circuit implementations for 4-bit reversible 
functions.  These algorithms have a number of potential uses and implications.

One major implication of this work is that it will help physicists with experimental design, since 
fore-knowledge of the optimal circuit implementation aids in the control over quantum mechanical systems.
The control of quantum mechanical systems is very difficult, and as a result experimentalists are always 
looking for the best possible implementation. Having an optimal implementation helps to improve experiments 
or show that more control over a physical system needs to be established before a certain experiment could be 
performed. To use our results in practice requires defining minimization criteria 
(e.g., implementation cost of gates VS depth VS architecture, etc.) dictated by a particular technology used, 
that may differ from one quantum information processing approach to another. Consequently, 
in this paper, we ignored such physical constraints, but concentrated on the minimization of the gate count.  
This serves as a proof of principle, showing that the search is possible in any practical scenario.  
We further explain how to modify our algorithms to account for more complex circuit minimization criteria 
in Section \ref{s:cfr}, and illustrate one of such modifications in the Section \ref{Extension}.

A second important contribution is due to the efficiency of our implementation---$.00756$ seconds per 
synthesis of an optimal 4-bit reversible circuit.  The algorithm could easily be integrated as part of 
peephole optimization, such as the one presented in \cite{ar:pspmh}.

Furthermore, our implementation allows to develop a subset of optimal implementations that may be used 
to test heuristic synthesis algorithms. Currently, similar tests are performed by comparison to optimal 
3-bit implementations \cite{ar:gaj, co:k, ar:mdm}. The best heuristic solutions have very tiny overhead when compared to optimal implementations, 
making such a test hard to improve. As such, it would help to replace this test with a more difficult one that 
allows more room for improvement. We suggest that this test set should include known benchmarks, and a 
combination of other functions---linear reversible, as well as, possibly, representatives from other classes, those with 
few gates and those requiring a large number of gates, etc.  We have not worked out the details of such a test. 

Finally, due to the effectiveness of our approach, we are able to report new optimal implementations for small 
benchmark functions, calculate $L(4)$, the number of reversible gates required to implement a reversible 4-bit function, 
calculate the average number of gates required to implement a 4-bit permutation, and show the distribution of the number 
of permutations that may be implemented with the predefined number of gates.

An earlier version of this paper has been presented at the DAC'2010 conference.

\section{Preliminaries}

\subsection{Quantum Computing}
We start with a very short review of basic concepts in quantum computing.  An in-depth coverage may be found in \cite{bk:nc}.

The state of a single qubit is described by a linear combination (Dirac notation)/column vector $\alpha |0\rangle + \beta |1\rangle = (\alpha, \beta)^t$, 
where $\alpha$ and $\beta$ are complex numbers called the amplitudes, and
$|\alpha|^2+|\beta|^2=1$. Real numbers $|\alpha|^2$ and $|\beta|^2$
represent the probabilities of reading the logic
states $|0\rangle$ and $|1\rangle$ upon (computational basis) measurement.  The state of a
quantum system with $n$ qubits is described by an element of the tensor
product of the single state spaces and can be represented as a
normalized vector of length $2^n$, called the state vector.  Furthermore,
quantum mechanics allows evolution of the state vector through its
multiplication by $2^n \times 2^n$ unitary matrices called the
gates.  These gates may be applied to a quantum state sequentially---such process constitutes constructing a 
circuit---which is equivalent to a series of proper matrix multiplications. 
To illustrate the gate application, take the two qubit state vector $|11\rangle = (0,0,0,1)^t$ 
and apply a CNOT gate, defined as the matrix
\[
\left[ {\begin{array}{*{20}c}
   1 & 0 & 0 & 0  \\
   0 & 1 & 0 & 0  \\
   0 & 0 & 0 & 1  \\
   0 & 0 & 1 & 0  \\
\end{array}} \right].
\]
The result is the state $|10\rangle = (0,0,1,0)^t$.  It may be observed 
that in the Dirac notation the CNOT gate may be described as follows: application 
of the CNOT gate flips the value of the second qubit iff the value of the first qubit is one.  
Dropping the bra-ket Dirac notation results in the following re-definition over 
Boolean values $a$ and $b$---gate CNOT performs transformation $a,b\mapsto a, b\oplus a$. This definition extends  
to vectors by the linearity, and thus is not ambiguous.  In particular, in the follow up 
sections we will consider reversible circuits (sometimes known as quantum Boolean 
circuits)---those where the matrix entries are strictly Boolean/integer, and for simplicity 
we will drop bra and ket in the notations, leaving just the variable names. The 
set of reversible circuits forms a group, that is also a subgroup of the set of all unitary 
transformations.

\subsection{Reversible Circuits}

\begin{figure}[t]
\centering
\includegraphics[height=22mm]{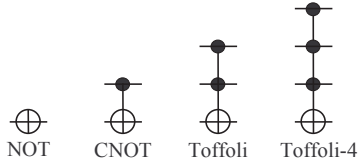}
\caption{NOT, CNOT, Toffoli, and Toffoli-4 gates.}
\label{fig1}
\end{figure}

In this paper, we consider circuits with NOT, CNOT, Toffoli (TOF), and Toffoli-4 (TOF4) gates defined as follows:
\begin{itemize}
\item NOT$(a):\; a\mapsto a\oplus 1$;
\item CNOT$(a,b):\; a,b\mapsto a, b\oplus a$;
\item TOF$(a,b,c):\; a,b,c\mapsto a, b, c\oplus ab$;
\item TOF4$(a,b,c,d):\; a,b,c,d\mapsto a, b, c, d\oplus abc$;
\end{itemize}
\noindent where $\oplus$ denotes the EXOR operation and concatenation is the Boolean AND; see Figure \ref{fig1} for illustration. 
These gates are used widely in quantum circuit construction, and have been demonstrated experimentally in multiple quantum information processing
proposals \cite{bk:nc}. In particular, CNOT is a very popular gate among experimentalists, frequently used to demonstrate 
control over a multiple-qubit quantum mechanical system. Since quantum circuits describe time evolution of a quantum 
mechanical system where individual ``wires'' represent physical instances, and time propagates from left to right, this imposes 
restrictions on the circuit topology. In particular, quantum and reversible circuits are strings of gates. As a result, 
feed-back (time wrap) is not allowed and there may be no fan-out (mass/energy conservation).

In this paper, we are concerned with searching for circuits requiring a minimal number of gates. Our focus is on the proof 
of principle, i.e., showing that any optimal 4-bit reversible function may be synthesized efficiently, rather than attempting 
to report optimal implementations for a number of potentially plausible cost metrics. In fact, our implementation allows other 
circuit cost metrics to be considered, as discussed in Section \ref{s:cfr} and Section \ref{Extension}.

In related work, there have been a few attempts to synthesize optimal reversible circuits with more than three inputs. 
Gro{\ss}e {\it et al.} \cite{co:gcdd} employ SAT-based technique to synthesize provably optimal circuits for some small parameters.
However, their implementation quickly runs out of resources. The longest optimal circuit they report contains 11 gates. 
The latter took 21,897.3 seconds to synthesize---same function that the implementation we report in this paper 
synthesized in .000052 seconds, see Table \ref{tab3}.
Prasad {\it et al.} \cite{ar:pspmh} used breadth first search to synthesize 26,000,000 optimal 4-bit reversible circuits 
with up to 6 gates in 152 seconds. We extend this search into finding all $16!$ optimal circuits in 1,130,276 seconds. 
This is over 100 times faster (per circuit) and 800,000 times more than reported in \cite{ar:pspmh}.
Yang {\it et al.} \cite{ar:yshp} considered short optimal reversible 4-bit circuits composed with NOT, CNOT, and 
Peres \cite{ar:p} gates. They were able to synthesize optimal circuits with up to 6 gates, and use those 
to optimally synthesize any given even permutation requiring no more than 12 gates. 
In other words, they can search a space of the size equal to approximately one quarter of the number of all 4-bit reversible functions. Our algorithms and  
implementation allow optimal synthesis of {\em all} 4-bit reversible functions and {\em any} 4-bit reversible function, and it is much faster. 

\subsection{Motivating Example}
\begin{figure}[t]
\centering
\includegraphics[height=22mm]{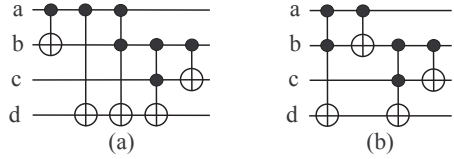}
\caption{(a) a suboptimal and (b) an optimal circuit for 1-bit full adder.}
\label{fig}
\end{figure}

Consider the two reversible circuit implementations in Figure~\ref{fig} of a 1-bit full adder. This elementary 
function/circuit serves as a building block for constructing integer adders. The famous Shor's integer factoring 
algorithm is dominated by adders like this.  As such, the complexity of an elementary 1-bit adder circuit largely 
affects the efficiency of factoring an integer number with a quantum algorithm. It is thus important to have a 
well-optimized implementation of a 1-bit adder, as well as other similar small quantum circuit building blocks. 

In this paper, we consider the synthesis of optimal circuits, i.e., we provably find the best possible implementation. 
Using optimal implementations of circuits potentially increases the efficiency of quantum algorithms and helps 
to reduce the difficulty with controlling quantum experiments.

\section{FINDOPT: an Algorithm to Find {\em an} Optimal Circuit}

We first outline our algorithm for finding an optimal circuit 
and then discuss it in detail in the follow up subsections.

There are $N = 2^n!$ reversible $n$-variable functions. The most obvious
approach to the synthesis of all optimal implementations is to compute all
optimal circuits and store them for later look-up. However, this is extremely inefficient.  This is because such an
approach requires $\Omega(N)$ space and, as a result, at least $\Omega(N)$ time. These space and
time estimates are lower bounds, because, for instance, storing an optimal circuit requires more
than a constant number of bits, but for simplicity, let us assume these figures are exact.
Despite considering both figures for space and time impractical,
we use this simple idea as our starting point.

We first improve the space
requirement by observing that if one synthesized all halves of all optimal circuits,
then it is possible to search through this set to find both halves of any optimal
circuit. It can be shown that the space requirement for storing halves has a lower bound of
$\Omega(\sqrt{N})$. However, searching for two halves potentially requires a runtime
on the order of the square of the search space, $\Omega\left((\sqrt{N})^2\right)=\Omega(N)$,
a figure for runtime that we deemed inefficient. Our second improvement is
thus to use a hash table to store the optimal halves. This reduces the runtime
to soft $\Omega(\sqrt{N})$. While this lower bound does not necessarily imply
that the actual complexity is lower than $O(N)$, this turns out to be the case,
because the set of optimal halves is indeed much smaller than the set of all 
optimal circuits (an analytic estimate for the relative size of the former set 
is hard to obtain, though). Cumulatively, these two improvements reduce $\Omega(N)$ 
space and $\Omega(N)$ time requirement to $O(\#{\rm halves}(N))$ space and
soft $O(\#{\rm halves}(N))$ time requirement. These reductions almost suffice
to make the search possible using modern computers.

Our last step, apart from careful coding, that made the search possible is the
reduction of the space requirement (with consequent improvement for runtime) by a
constant of almost 48 via exploiting the following two features. First, simultaneous
input/output relabeling, of which there are at most 24 (=4!) different ones, does not change the
optimality of a circuit. And second, if an optimal circuit is found for a function $f$, an optimal circuit for the inverse function, $f^{-1}$, 
can be obtained by reversing the optimal circuit for $f$. This allows to additionally ``pack'' up
to twice as many functions into one circuit. The cumulative improvement
resulting from these two observations, is by a factor of almost $2\times 24=48$.
Due to symmetries, the actual number is slightly less. See Table \ref{tab1}
(column 2 versus column 3) for exact comparison.

\subsection{The search-and-lookup algorithm}

For brevity, let the
size of a reversible function mean the minimal number of gates required
to implement it. Using breadth-first search, we can generate
the smallest circuits for all reversible functions of size at most $k$,
for a certain value of $k$.
(This can be done in advance, on a larger machine, and need not be repeated
for each reversible function.)

Assume that the given function $f$, for which we need to synthesize
a minimal circuit, has size at most $2k$. We can first check whether
$f$ is among the known functions of size at most $k$ and, if so,
output the corresponding minimal circuit. If not, then the size of
$f$ is between $k+1$ and $2k$, inclusive, and there exist reversible
functions $h$ and $g$ of size $k$ and at most $k$, respectively, such
that $f=h\circ g$. If we find such $g$ of the smallest size, then we
can obtain the smallest circuit for $f$ by composing the circuits for
$g$ and $h$.

Multiplying the above equality by $g^{-1}$, we obtain $f\circ g^{-1}=h$.
Observe that $g^{-1}$ has the same size as $g$. Therefore, by trying
all functions $g$ of size $1,2,\ldots,k$ until we find one such
that $f\circ g$ has size $k$, we can find a $g$ of the smallest size.

The above algorithm involves sequential access to the functions of size
at most $k$ and their minimal circuits and a membership test among
functions of size $k$. Since the latter test must be fast and requires
random memory access, we need to store all functions of size $k$ in the memory.
Thus, the amount of available RAM imposes an upper bound on $k$.

In practice, we store a 4-bit reversible function using a 64-bit word,
because this allows for an efficient implementation of functional
composition, inversion, and other necessary operations.
On a typical PC with 4GB of RAM, we can store all functions for $k=6$.
This means that we can apply the above search algorithm only to functions
of size at most 12. Unfortunately, this will not cover all 4-bit reversible
functions. Therefore, further reduction of memory usage is necessary.

\subsection{Symmetries}

A significant reduction of the search space can be achieved by taking into
account the following symmetries of circuits:

\begin{enumerate}
  \item Simultaneous relabeling of inputs and outputs. Given an optimal circuit
implementing a 4-bit reversible function $f$ with inputs $x_0,x_1,x_2,x_3$
and outputs $y_0,y_1,y_2,y_3$ and a permutation $\sigma:\{0,1,2,3\}\to\{0,1,2,3\},$
we can construct a new circuit by relabeling the inputs and outputs into
$x_{\sigma(0)},x_{\sigma(1)},x_{\sigma(2)},x_{\sigma(3)}\;{\rm and}\;
y_{\sigma(0)},y_{\sigma(1)},y_{\sigma(2)},y_{\sigma(3)},$ respectively.
Then the new circuit will provide a minimal implementation of the corresponding
reversible function $f_\sigma$. Indeed, if it is not minimal and
there is an implementation of $f_\sigma$ by a circuit with a smaller number of gates,
we can relabel the inputs and outputs of this implementation with $\sigma^{-1}$
and obtain a smaller circuit implementing the original function $f$. This contradicts
the assumption that the original circuit for $f$ is optimal.

Given $f$ and $\sigma$, a formula for $f_\sigma$ can be easily obtained.
Observe that the mapping
$x_0,x_1,x_2,x_3 \mapsto x_{\sigma(0)},x_{\sigma(1)},x_{\sigma(2)},$ $x_{\sigma(3)}$
is a 4-bit reversible function, which we denote by $\tilde\sigma$. The mapping
$y_{\sigma(0)},y_{\sigma(1)},y_{\sigma(2)},y_{\sigma(3)} \mapsto y_0,y_1,y_2,y_3$
is then given by the inverse, $\tilde\sigma^{-1}$. Therefore, the four bit values
$y_0,y_1,y_2,y_3$ of $f_\sigma$ on a four-bit tuple $x_0,x_1,x_2,x_3$ can be
 obtained by applying first $\tilde\sigma$, then $f$, and finally $\tilde\sigma^{-1}$.
We obtain
$f_\sigma = \tilde\sigma^{-1} \circ f \circ \tilde\sigma.$
We call the set of functions $f_\sigma$ the {\it conjugacy class} of $f$
modulo simultaneous input/output relabelings.

Since there exist 24 permutations of 4 numbers, by choosing different permutations
$\sigma$, we obtain 24 functions of the above form $f_\sigma$ for a fixed function $f$.
Some of these functions may be equal, whence the size of the conjugacy class of $f$
may be smaller than 24. For example, if $f$=NOT$(a)$, then there exist only $4$ distinct
functions of the form $f_\sigma$ (counting $f$ itself). Our experiments show, however,
that for the vast majority of functions, the conjugacy classes are of size 24.

  \item Inversion. As mentioned above, if we know a minimal implementation for $f$,
then we know one for its inverse as well.
\end{enumerate}

Note that conjugation and inversion commute:
$$(\tilde\sigma^{-1}\circ f \circ \tilde\sigma)^{-1} = \tilde\sigma^{-1}\circ f^{-1} \circ \tilde\sigma.$$
For a function $f$, consider the union of the two conjugacy classes of $f$ and $f^{-1}$.
Call the elements of this union {\it equivalent} to $f$. It follows that equivalent
functions have the same size. Moreover, since gates are idempotent (i.e., equal
to their own inverses) and their conjugacy classes consist of gates, if we know a
minimal circuit for $f$, we can easily obtain one for any function in the equivalence class of $f$.
Formally, if $f=\lambda_1\circ\ldots\circ \lambda_n$, where $n$ is the size of $f$ and
$\lambda_i$ are gates, then $f^{-1} = \lambda_n\circ\ldots\circ \lambda_1$, and if
$f'=\tilde\sigma^{-1}\circ f\circ \tilde\sigma$, then
$f'=\lambda_1'\circ\ldots\circ \lambda_n'$, where $\lambda_i' = \tilde\sigma^{-1}\circ \lambda_i\circ \tilde\sigma$
are also gates.
Our experiments show that a vast majority of functions have 48 distinct equivalent
functions. This fact can reduce the search space by almost a factor of 48 as follows.

For a function $f$, define the canonical representative of its equivalence class.
A convenient canonical representative can be obtained by introducing the lexicographic
order on the set of 4-bit reversible functions, considered as permutations of
$\{0,1,2,\ldots,15\}$ and encoded accordingly by the sequence $f(0),f(1),\ldots,f(15)$,
and choosing the function whose corresponding sequence is lexicographically smallest.
Now, instead of storing all functions of size at most $k$, store the canonical
representative for each equivalence class. This will reduce the storage size
by almost a factor of 48.
Then, we use Algorithm~\ref{slsym} to search for a minimal circuit for a
given reversible function $f$.

\begin{algorithm}[t]
\caption{Minimal circuit (FINDOPT).}
\label{slsym}
\begin{algorithmic}
\REQUIRE   Reversible function $f$ of size at most $L$.\\
 Hash table $H$ containing canonical representatives of all equivalence
classes of functions of size at most $k$ and the last gates of their minimal
circuits, $k\ge L/2$.\\
 Lists $A_i$, $1\le i\le L-k$, of all functions of size $i$.
\ENSURE A minimal circuit $c$ for $f$.

\IF{$f=\;$IDENTITY}
  \RETURN{empty circuit}
\ENDIF

\STATE $E_f\leftarrow $ equivalence class of $f$
\STATE $\bar f\leftarrow$ canonical representative of $E_f$
\IF{$\bar f\in H$}
  \STATE $\bar\lambda\leftarrow$ last gate of $\bar f$
  \IF{$f$ is a conjugate of $\bar f$}
    \STATE let $f = \tilde\sigma^{-1}\circ \bar f \circ \tilde\sigma$
    \STATE $\lambda\leftarrow \tilde\sigma^{-1}\circ\bar\lambda \circ \tilde\sigma$
    \STATE $c\leftarrow$ minimal circuit for $f\circ\lambda$
    \RETURN {$c\circ \lambda$}
  \ELSE
    \STATE let $f = \tilde\sigma^{-1}\circ \bar f^{-1} \circ \tilde\sigma$
    \STATE $\lambda\leftarrow \tilde\sigma^{-1}\circ\bar\lambda \circ \tilde\sigma$
    \STATE $c\leftarrow$ minimal circuit for $\lambda \circ f$
    \RETURN {$\lambda\circ c$}
  \ENDIF
\ENDIF

\FOR{$i=1$ to $L-k$}
  \FOR {$g\in A_i$}
    \STATE $h\leftarrow g\circ f$
    \STATE $E_h\leftarrow$ equivalence class of $h$
    \STATE $\bar h\leftarrow$ canonical representative of $E_h$
    \IF{$\bar h\in H$}
       \STATE $c_g\leftarrow$ minimal circuit for $g$
       \STATE $c_h\leftarrow$ minimal circuit for $h$
       \RETURN{$c_g^{-1}\circ c_h$}
    \ENDIF
  \ENDFOR
\ENDFOR

\RETURN{{\bf error:} size of $f$ is greater than $L$}

\end{algorithmic}

\end{algorithm}

The algorithm requires a hash table with canonical representatives of equivalence classes
of size at most $k$, together with the last gates of their minimal circuits, and lists
of all permutations of size at most $L-k$.
We have pre-computed the canonical representatives for $k=9$ using breadth-first search
(see Algorithm~\ref{bfs}).
For efficiency reasons, we store the last {\it or the first} gate of
a minimal circuit for each canonical representative. However, this information is clearly
sufficient to reconstruct the entire circuit and, in particular, the last gate.
Using this pre-computed data, the hash table and the lists of all permutations of size
at most $L-k$ are formed at the start-up. An implementation storing only the hash table is
possible. Such an implementation will require less RAM memory, but it will be slower. We
decided to focus on higher speed, because Table \ref{tab1} indicates that we do not
need to be able to search optimal circuits requiring up to 18 ($=9\times 2$) gates,
which we could do otherwise by storing only the hash table.

\begin{algorithm}[t]
\caption{Breadth-first search (BFS).}
\label{bfs}
\begin{algorithmic}
  \REQUIRE $k$
  \ENSURE Lists $A_i$ of canonical representatives of size $\le k$;\\
    Hash table $H$ with these canonical representatives and their first or last gates.\\
  \STATE Let $H$ be a hash table (keys are functions, values are gates)
  \STATE $H$.insert(IDENTITY, HAS\_NO\_GATES)
  \STATE $A_0\leftarrow\{{\rm IDENTITY}\}$
  \FOR {$i$ from $1$ to $k$}
    \FOR {$f\in A_{i-1} \cup \{a^{-1}\;|\;a\in A_{i-1}\}$}
      \FOR {all gates $\lambda$}
         \STATE $h\leftarrow f\circ\lambda$
         \STATE $E_h\leftarrow$ equivalence class of $h$
         \STATE $\bar h\leftarrow$ canonical representative of $E_h$
         \IF {$\bar h\not\in H$}
            \IF {$h$ is a conjugate of $\bar h$}
              \STATE let $h=\tilde\sigma^{-1}\circ \bar h\circ \tilde\sigma$
              \STATE $H$.insert($\bar h$, $\tilde\sigma^{-1}\circ \lambda\circ \tilde\sigma$, IS\_LAST\_GATE)
            \ELSE
              \STATE let $h=\tilde\sigma^{-1}\circ \bar h^{-1}\circ \tilde\sigma$
              \STATE $H$.insert($\bar h$, $\tilde\sigma^{-1}\circ \lambda\circ \tilde\sigma$, IS\_FIRST\_GATE)
            \ENDIF
            \STATE $A_i$.insert($\bar h$)
         \ENDIF
       \ENDFOR
     \ENDFOR
  \ENDFOR
\end{algorithmic}
\end{algorithm}

The correctness of Algorithm~\ref{slsym} is proved as follows. Suppose first that
the size of $f$ is at most $k$. The canonical representative $\bar f$
of its equivalence class will have the same size as $f$, so it will be
found in the hash table $H$. Since $\bar\lambda$ is the last gate of a
minimal circuit for $\bar f$, the size of $\bar f\circ\bar \lambda$ is
one less than the size of $\bar f$. The function $f\circ\lambda$
(computed if $f$ is a conjugate of $\bar f$) or
the function $\lambda\circ f$ (computed if $f$ is a conjugate of $\bar f^{-1}$)
is equivalent to $\bar f\circ\bar\lambda$ and therefore also is of size one
less than the size of $\bar f$. Therefore, the recursive call on that function
will terminate and return a minimal circuit, which we can compose with $\lambda$
(at the proper side) to obtain a minimal circuit for $f$. The depth of recursion
is equal to the size of $f$, and at each call we do one hash table lookup,
one computation of the canonical representative, and one conjugation of a gate
(the latter can be looked up in a small table). Thus, this part
of the algorithm requires negligible time.

If the size of $f$ is greater than $k$, but does not exceed $L$, then
$f=g_f\circ h$ for some $h$ of size $k$ and $g_f$ of size $i$, $1\le i\le L-k$.
Then $g=g_f^{-1}\in A_i$. Once the inner for-loop encounters this $g$, it
will return the minimal circuit for $f$, because both recursive calls are
for functions of size at most $k$. For a function $f$ of size $s>k$,
the number of iterations required to find the minimal circuit satisfies
$$\sum_{i=1}^{s-1-k}|A_i|<r\le \sum_{i=1}^{s-k}|A_i|.$$
At each iteration, one canonical representative is computed and looked up
in the hash table. Since the size of $A_i$ grows almost exponentially
(see Table~\ref{tab1}, left column), the search time will decrease
almost exponentially, and the storage will increase exponentially,
as $k$ increases. The timings for $k=8,9$ measured on two different
systems are summarized in Table~\ref{timings} (see Section~\ref{sec:performance} for machine details).  Please, note 
that size 15 circuits may be verified against Table \ref{tab:15shki} and consequently the time to synthesize them, 
for all practical purposes, is zero.  We marked relevant entries in the Table~\ref{timings} with an asterisk.  
The hash table loading and overall memory usage times were 191 seconds, 3.5GB ($k=8$) and
1667 seconds, 43.04GB ($k=9$).

\begin{table}
\begin{center}
\caption{Average times of computing minimal circuits of sizes $0..15$ (in seconds).}
\label{timings}
\begin{tabular}{|l|rrr|}
\hline
Size \hfill $\setminus$ \hfill $k$ & 8 (LPTP) & 8 (CLSTR) & 9 (CLSTR)\\
\hline
1 & $8.70\times 10^{-7}$ & $5.25\times 10^{-7}$ & $5.23\times 10^{-7}$ \\
2 & $1.26\times 10^{-6}$ & $8.32\times 10^{-7}$ & $8.33\times 10^{-7}$ \\
3 & $1.66\times 10^{-6}$ & $1.14\times 10^{-6}$ & $1.15\times 10^{-6}$ \\
4 & $2.07\times 10^{-6}$ & $1.47\times 10^{-6}$ & $1.47\times 10^{-6}$ \\
5 & $2.47\times 10^{-6}$ & $1.79\times 10^{-6}$ & $1.79\times 10^{-6}$ \\
6 & $3.48\times 10^{-6}$ & $2.11\times 10^{-6}$ & $2.12\times 10^{-6}$ \\
7 & $4.22\times 10^{-6}$ & $2.46\times 10^{-6}$ & $2.46\times 10^{-6}$ \\
8 & $4.49\times 10^{-6}$ & $2.81\times 10^{-6}$ & $2.80\times 10^{-6}$ \\
9 & $1.07\times 10^{-5}$ & $6.68\times 10^{-6}$ & $3.11\times 10^{-6}$ \\
10& $2.28\times 10^{-4}$ & $9.31\times 10^{-5}$ & $6.23\times 10^{-6}$ \\
11& $4.27\times 10^{-3}$ & $3.60\times 10^{-3}$ & $7.23\times 10^{-5}$ \\
12& $6.30\times 10^{-2}$ & $5.58\times 10^{-2}$ & $1.34\times 10^{-3}$ \\
13& $4.91\times 10^{-1}$ & $4.80\times 10^{-1}$ & $2.20\times 10^{-2}$ \\
14& $4.38\times 10^0\;\,\,$ & $4.50\times 10^0\;\,\,$    & $2.32\times 10^{-1}$ \\
15& N/A$^{*}$          & $6.14\times 10^{1*}\;$    & $3.61\times 10^{0*}\;$ \\
\hline
\end{tabular}
\end{center}
\end{table}

It follows from the above complexity analysis that the performance of the
following key operations affect the speed most:
\begin{itemize}
  \item composition of two functions ($f\circ g$) and inverse of a function ($f^{-1}$),
  \item computation of the canonical representative of an equivalence class,
  \item hash table lookup.
\end{itemize}
In the next Subsection we discuss an efficient implementation of these operations.

\subsection{Implementation details}

As mentioned above, a 4-bit reversible function can be stored in
a 64-bit word, by allocating 4 bits for each value of
$f(0),f(1),\ldots,f(15)$.
Then the composition of two functions can be computed in 94 machine
instructions using the algorithm {\tt composition} and
the inverse function can be computed in 59 machine instructions
using algorithm {\tt inverse}.

\begin{figure}[h]
\begin{small}
\begin{verbatim}
unsigned64 composition(unsigned64 p,
                       unsigned64 q) {
  unsigned64 d = (p & 15) << 2;
  unsigned64 r = (q >> p_i) & 15;
  p >>= 2; d = p & 60;
  r |= ((q >> d) & 15) << 4;
  p >>= 4; d = p & 60;
  r |= ((q >> d) & 15) << 8;
  p >>= 4; d = p & 60;
  r |= ((q >> d) & 15) << 16;
  ...
  p >>= 4; d = p & 60;
  r |= ((q >> d) & 15) << 60;
  return r;
}

unsigned64 inverse(unsigned64 p) {
  p >>= 2;
  unsigned64 q =  1 << (p & 60);
  p >>= 4;  q |=  2 << (p & 60);
  p >>= 4;  q |=  3 << (p & 60);
  ...
  p >>= 4;  q |= 15 << (p & 60);
  return q;
}

unsigned64 conjugate01(unsigned64 p) {
  p = (p & 0xF00FF00FF00FF00F)        |
      ((p & 0x00F000F000F000F0) << 4) |
      ((p & 0x0F000F000F000F00) >> 4);
  return (p & 0xCCCCCCCCCCCCCCCC)        |
         ((p & 0x1111111111111111) << 1) |
         ((p & 0x2222222222222222) >> 1);
}
\end{verbatim}
\end{small}
\end{figure}

In order to find the canonical representative in the equivalence class
of a function $f$, we compute $f^{-1}$, generate all conjugates
of $f$ and $f^{-1}$, and choose the smallest among the resulting 48
functions. Since every permutation of $\{0,1,2,3\}$ can be represented
as a product of transpositions $(0,1)$, $(1,2)$, and $(2,3)$,
the sequence of conjugates of $f$ by all 24 permutations can be
obtained through conjugating $f$ by these transpositions. These
conjugations can be performed in 14 machine instructions each
as in function {\tt conjugate01}.

Two functions can be compared lexicographically using a single
unsigned comparison of the corresponding two words. Thus, the canonical
representative can be computed using one inversion, $23\times 2 = 46$
conjugations by transpositions, and 47 comparisons, which totals
to 750 machine instructions.

For the fast membership test, we use a linear probing
hash table with Thomas Wang's hash function \cite{wang} (see algorithm {\tt hash64shift}).

\begin{figure}[h]
\begin{small}
\begin{verbatim}
long hash64shift(long key) {
  key = (~key) + (key << 21);
  key = key ^ (key >>> 24);
  key = (key + (key << 3)) + (key << 8);
  key = key ^ (key >>> 14);
  key = (key + (key << 2)) + (key << 4);
  key = key ^ (key >>> 28);
  key = key + (key << 31);
  return key;
}
\end{verbatim}
\end{small}
\end{figure}

This function is well suited for our purposes: it is fast
to compute and distributes the permutations uniformly over the
hash table. The parameters of the hash tables storing
the canonical representatives of equivalence classes of size $k$,
for $k=7,8,9$ are shown in Table~\ref{hashtables}.

\begin{table}
\begin{center}
\caption{Parameters of linear hash tables storing canonical representatives.}
\label{hashtables}
\begin{tabular}{|l|ccc|}
\hline
\hfill $k$ & 7 & 8 & 9 \\
\hline 
Size$\vphantom{2^{25^{25}}}$ & $2^{25}$ & $2^{28}$ & $2^{32}$ \\
Memory Usage & 256 MB & 2 GB & 32 GB\\
Load Factor & 0.58 & 0.84 & 0.51 \\
Average Chain Length & 3.14 & 9.18 & 2.63\\
Maximal Chain Length & 92 & 754 & 86\\
\hline
\end{tabular}
\end{center}
\end{table}

\section{SEARCHALL: an Algorithm to Find {\em all} Optimal Circuits}

We first outline our algorithm for finding all optimal circuits 
and then discuss it in detail in the follow up subsections.

We employ a breadth first search that consists of two stages:
\begin{itemize}
\item Optimal circuits with 0..9 gates are found with Algorithm~\ref{bfs}, BFS. 
This algorithm becomes inefficient for finding optimal circuits with 10 or more gates.
\item Optimal circuits with 10 and more gates are found by storing and updating 
the bit vector of canonical representatives of permutations requiring a certain 
number of gates.
\end{itemize}

The SEARCHALL algorithm is used to find all reversible functions of size $k$
for $k=10,11,\ldots$, until we reach the maximal size of a reversible function.
Starting from the known set of reversible functions of size 9, we consecutively
proceed to sizes $10,11,\ldots$. The transitions from size $k$ to size $(k+1)$
are carried out as follows (Subsections \ref{ss:a} to \ref{ss:d}). 

First, we choose a compact representation for the set of reversible functions
of size $k$, based on the following concept of an almost reduced function.

\subsection{Almost reduced functions} \label{ss:a}

Call a reversible function (permutation) $p$ {\it almost reduced} if one of the
following two conditions holds:
\begin{enumerate}
  \item $p(0)=0$ and $p(15)\in\{1,3,7,15\}$
  \item $(p(0),p^{-1}(0))$ belongs to the following set
    $$\begin{array}{ll}
     \{&(1,1),(1,2),(1,15),(3,1),(3,3),(3,4),\\
       &(3,5),(3,12),(3,15),(7,1),(7,3),(7,7),\\
       &(7,8),(7,9),(7,11),(7,15),(15,15)\quad\}\end{array}$$
\end{enumerate}

\begin{lemma}
  For every permutation $p$, there is at least one equivalent
  almost reduced reversible function.
\end{lemma}

Note that a reduced reversible function is not necessarily almost reduced. This
will hopefully not lead to a confusion, since we are not going to deal with
reduced functions in this section.

An almost reduced permutation $p$ can be uniquely specified by the following data:
\begin{itemize}
  \item $A_p$: $p(0)$
  \item $B_p$: $p(15)$ if $p(0)=0$, otherwise $p^{-1}(0)$
  \item $Q_p$: a permutation of 14 elements.
\end{itemize}
We call this data an {\it indexable specification} of the almost reduced permutation $p$.

The set of almost reduced reversible functions can be totally ordered by ordering their
indexable specifications $(A_p,B_p,Q_p)$ lexicographically. The {\it index} of
an almost reduced reversible function $p$ is defined as the number of smaller almost
reversible functions. In order to compute the index of a function $p$, we first
compute its indexable specification $(A_p,B_p,Q_p)$. Then we compute $n(Q_p)$,
the number of 14-element permutations smaller than $Q_p$, and $n(A_p,B_p)$, the number
of pairs $(A,B)$ that are a valid part of an indexable specification and 
are lexicographically smaller than $(A_p,B_p)$. Then the index of $p$ is given by
$$n(p) = 14! n(A_p,B_p) + n(Q_p).$$

Efficient conversions between reversible functions and their indexable specifications
are quite straightforward, therefore we omit these algorithms here. Various efficient
algorithms for indexing permutations are also well-known.

Since almost reduced functions and their indexable specifications are in a one-to-one
correspondence, the total number of almost reduced functions is
 $21*14! < 1.84\times 10^{12}$. 
This is $\approx 11.43$ times less than the total number of reversible functions, yet 
about 4 times greater than the number of equivalence classes---i.e., different reduced permutations. The main reason why
we do not index equivalence classes directly (which would have further reduced our memory
requirements by about a factor of 4) is that we could not find an efficient
algorithm for computing these indices.

\subsection{From size $k$ to size $k+1$}

We encode the set of reversible functions of size $k$ by a bit array of size $21\times 14!$ bits 
(under 209GB), where bit $i$ is set whenever the almost reduced function $p$ with $n(p)=i$
has size $k$ and is the smallest almost reduced function in its equivalence class.

We further split this array in 3 parts called {\it slices}, by partitioning the set of pairs $(A,B)$ that
are valid parts of an indexable specification into 3 subsets. One third of the bit
array easily fits in the memory of the machine we were using for the experiments
(and leaves enough extra space for the system not to be tempted to turn on swapping
 during the computation).

Suppose that the bit array for functions of size $k$ is stored in an input file. We compute
the bit array for functions of size $k+1$ and store it in the output file via the
following stages:
\begin{enumerate}
  \item Composition. Repeated for each target slice $s$ (there are three of them).
 Allocate in memory a bit array $a$ of size $21\times 14! / 3$ bits.
 For every almost reduced function $p$ marked in the input bit array, generate all its
 conjugates and inverses (thus we obtain all reversible functions of size $k$). 
 Then for every function $p'$ in the equivalence class of $p$
 and every gate $g$, find an almost reduced representative $q$ in the equivalence class
 of the composition $g\circ p'$, then compute its index $n(q)$. If $n(q)$ is in slice $s$,
 set the $n(q)$-th bit in the bit array $a$. At the end, output the array in a new file and
concatenate the three slices.
  \item Canonization. Because an equivalence class can have more than one almost
reduced element, the previous stage may have marked more than one bit for some
equivalence classes $g\circ p$. We scan the bit array output at the previous stage and,
for each permutation $q$ marked there, compute the smallest equivalent almost
reduced permutation $q'$ and mark the corresponding bit in the array allocated in
memory. Since the entire bit array does not fit in memory, we again use three slices and
at the end concatenate them.
  \item Subtraction. The bit array produced by the previous stage contains all functions
of size $k+1$, as well as some functions of size $k$ and $k-1$. We therefore subtract
the bit arrays corresponding to sizes $k$ and $k-1$. The resulting bit array satisfies
the property: bit $i$ is set whenever the almost reduced function $p$ with $n(p)=i$
has size $k+1$ and is the smallest almost reduced function in its equivalence class.
  \item (Optional stage) Counting. For each almost reduced function of size $k+1$, smallest
in its equivalence class, we generate the entire equivalence class and count its cardinality.
As a result, we obtain the total number of reversible functions of size $k+1$.
\end{enumerate}

\subsection{Optimization}

The hardest stage to optimize is Composition. Our initial implementation, which was quite
literally following the above description, with some {\it ad hoc} improvements, was going
to require months to compute the functions of size $12$. We found the following shortcut,
which speeds it up by about a factor of 24. 

For every almost reduced function $p$ marked in the input bit array, we compute its 
equivalence class. However, we avoid computing the compositions of each element of the
equivalence class and each gate. Instead, we extract the values
 $p(0),p(15),p^{-1}(0),p^{-1}(15)$. Given these values and a gate $g$, one can determine
which conjugation and possibly inversion must be applied to $g\circ p$ to obtain an almost
reduced function. The table of these conjugations and inversions is pre-computed in
advance.

Then, suppose a given permutation $p$ is conjugate to an almost reduced permutation, i.e.,
$c^{-1}\circ g\circ p\circ c$ for some conjugation $c$
is almost reduced. We rewrite this as $c^{-1}\circ g\circ c\circ c^{-1}\circ p\circ c$.
The conjugations of the 32 gates are also pre-computed in advance and stored in a separate table.
Since the conjugations of $p$ have been computed at the beginning of this step (indeed, we 
have computed the entire equivalence class of $p$), we can just take one of its
elements $p'$ and compose it with the gate $g'=c^{-1}\circ g\circ c$. The resulting
permutation $g'\circ p'$ is almost reduced.

If the almost reduced representative in the equivalence class of $g\circ p$ is a conjugate
of the inverse $(g\circ p)^{-1}$, i.e., equals $c^{-1}\circ (g\circ p)^{-1} \circ c$, then we rewrite
 this as $c^{-1}\circ p^{-1} \circ g^{-1}\circ c = c^{-1}\circ p^{-1}\circ c \circ c^{-1} \circ g\circ c$ 
(also using the fact that $g^{-1}=g$ for every gate $g$). Now we again observe that
$c^{-1}\circ p^{-1}\circ c$ has been pre-computed, so we only need to compose it
with the gate $g'=c^{-1}\circ g\circ c$. Note that compositions of functions with
gates (on either side) can be performed very efficiently.

Having implemented this optimization, we were able to compute all reversible functions of size $10$.  However, the computation of functions of size $12$ 
would still take too long, so we parallelized the algorithm.

\subsection{Parallelization} \label{ss:d}

Both composition and canonization stages are computationally intensive. We parallelized
them using the following architecture implemented with MPI.

For composition, the master job reads the input bit vector in blocks. Every block is sent to one of 16 
workers, which are chosen in a circular (round robin) order. These workers decode the bits in the blocks
 into permutations of size $k$, apply gates to them as described above, and compute the indices of the
 resulting almost reduced permutations. These indices are stored in a temporary array, which is partitioned
into 8 equal slices. Once all indices have been computed by a worker, the slices are sent to the
corresponding 8 collector jobs. Each collector possesses its own bit vector allocated
 in RAM. It receives arrays with indices of bits to be marked from the 16 workers in a round robin order.
Having received an array from a worker, it marks the corresponding bits.
At the end of a round, the collector signals the master that a round has been completed.
 At the very end, the collectors write their bit vectors to disk in sequence.

The master makes sure that the collectors are no more than 80 blocks behind it. If it continued to
send the blocks to the workers without waiting for the workers and collectors to finish processing them,
the unprocessed blocks would accumulate in the communication channels between the master and the
workers. This results in a memory leak, which turned out to be faster than the system swapping mechanism,
and therefore caused a deadlock. By allowing the collectors to be only a certain number of blocks behind
the master, we restrict the amount of data in the communication channels at any given moment and thus
 prevent the leak.
It is useful to allow a non-zero lag though, for otherwise the system becomes overly synchronized, which
drastically reduces the performance: the workers and collectors that finish first end up waiting on the
others most of the time. With the lag, communication channels work as buffers, from which the workers
continue to draw data. The amount of data in each particular channel at a given moment
may vary, depending on the speed at which the corresponding worker processes the previous blocks.

Exactly the same parallel architecture is used for canonization. The master again reads the input bit vector
in blocks. The workers compute the minimal almost reduced equivalent permutation for each almost reduced
permutation they receive from the master and send their indices to the collectors. The collectors mark the 
corresponding bits and write those bit vectors to disk at the end.

\section{Performance and Results}
\label{sec:performance}
We performed several tests using two computer systems, LPTP and CLSTR.
LPTP is a Sony VGN-NS190D laptop with
Intel Core Duo 2000 GHz processor, 4 GB RAM, and a 5400 RPM SATA HDD running Linux.
CLSTR is a cluster \cite{CLSTR} located at the Institute for Quantum Computing. 
We used a single Sun X4600 node with 128 GB RAM and 8 AMD Opteron quad-core CPUs for each run of the SEARCHALL and FINDOPT algorithm in CLSTR.
The following subsections summarize the tests and results.

\subsection{Synthesis of Random Permutations} \label{ss:1}

\begin{table}
\centering
\caption{Distribution of the number of gates required for 10,000,000 random 4-bit reversible functions.}
\label{tab2}
\begin{tabular}{|r|r|} \hline
Size  	& Functions 		\\  \hline
14		& 17,191 	\\
13		& 2,371,039 	\\
12		& 5,110,943		\\
11		& 2,051,507 		\\
10		& 392,108 		\\
9		& 50,861			\\
8		& 5,269 			\\
7		& 455 				\\
6		& 24				\\
5   	& 3         		\\ \hline
\end{tabular}
\end{table}

In this test, we generated 10,000,000 random uniformly distributed permutations using the Mersenne twister random 
number generator~\cite{ar:mn}. We next generated their optimal circuits using algorithm FINDOPT. The test was executed on CLSTR.
It took $75,613.12$ seconds (about $21$ hours) of user time and the
maximal RAM memory usage was 43.04GB. Note that 1667 seconds (approximately 28 minutes) were spent loading 
previously computed optimal circuits with up to 9 gates (see Subsection \ref{ss:2} for details) into RAM.
On average, it took only $0.00756$ seconds to synthesize an optimal circuit for a permutation. The distribution 
of the circuit sizes is shown in Table \ref{tab2}.

Since there are no permutations requiring 16 or more gates, and only a few permutations requiring 15 gates
(see Subsection \ref{ss:3} for details), this implies that the search FINDOPT 
may be easily modified to explicitly store all optimal 15-bit implementations in the cache, and search optimal implementations 
with up to 14 gates.  Such a search may be executed using a computer capable of storing reduced optimal implementations with up to 7 gates, 
i.e., a machine with only 256M of available RAM.  In other words, FINDOPT allows performing optimal 4-bit circuit calculation 
even on an older machine. 

\subsection{Distribution of Optimal Implementations} \label{ss:2}

\begin{table}[h]
\centering
\caption{Number of 4-bit permutations requiring prescribed number of gates.}
\label{tab1}
\begin{tabular}{|r|r|r|r|} \hline
Size  	& Functions 		 & Reduced         &  Runtime \\
      	&           		 & Functions       &          \\ \hline
$\geq$16& 0 				 & 0               &            \\
15		& 144 				 & 5 			   & 66,782s \\
14		& 37,481,795,636	 & 781,068,573     & 245,488s \\
13		& 4,959,760,623,552	 & 103,331,100,613 & 397,464s \\
12		& 10,690,104,057,901 & 222,714,352,278 & 238,589s \\
11		& 4,298,462,792,398	 & 89,554,073,333  & 103,595s \\
10		& 819,182,578,179	 & 17,067,688,249  & 68,670s \\
9		& 105,984,823,653 	 & 2,208,511,226   & 8,836.36s \\
8		& 10,804,681,959 	 & 225,242,556 	   & 744.41s \\ 
7		& 932,651,938 		 & 19,466,575 	   & 95.574s    \\ 
6		& 70,763,560 		 & 1,482,686 	   & 11.109s    \\ 
5		& 4,807,552 		 & 101,983 		   & 0.816s     \\ 
4		& 294,507 			 & 6,538 		   & 0.06s      \\ 
3		& 16,204 			 & 425 			   & 0.004s     \\
2		& 784 				 & 33			   & $<$0.001s  \\
1		& 32				 & 4			   & $<$0.001s  \\
0   	& 1     			 & 1			   & $<$0.001s  \\ \hline
Total   & 20,922,789,888,000 & 435,903,095,078 & 1,130,276s  \\ \hline
\end{tabular}
\end{table}

Table \ref{tab1} lists the distribution of the number of permutations that can be realized with 
optimal circuits requiring a specified number of gates.  We used CLSTR to run this test, and it took 1,130,276 seconds (approximately 13 days) to complete 
it.  
Circuits with up to 9 gates were synthesized using BFS algorithm.  For circuits with 10 gates and more 
we used SEARCHALL.

We have calculated the average number of gates required for a random 4-bit reversible function, $11.93937\ldots$.

\subsection{Most complex permutations} \label{ss:3}
\begin{table*}[t]
\centering
\caption{Permutations requiring 15 gates.}
\label{tab:15shki}
{\footnotesize
\begin{tabular}{|r|r|r|} \hline
Function  	& \# Symm. & Implementation   \\ \hline
[1,5,0,8,9,11,2,15,3,12,4,6,10,14,13,7]     & 24 & CNOT(a,c) CNOT(c,d) CNOT(d,a) TOF(b,d,c) CNOT(a,b) TOF(c,d,b) TOF4(a,b,c,d) \\
                                            &    & CNOT(c,a) NOT(b) NOT(c) CNOT(a,d) TOF(b,d,c) TOF(b,c,a) TOF(a,c,b) NOT(c) \\ \hline
[1,9,0,4,10,8,2,11,3,15,5,12,7,14,13,6]  & 24 & NOT(d) CNOT(d,c) TOF4(a,c,d,b) TOF(a,d,c) TOF(b,d,a) TOF(c,d,b) TOF(b,c,d) \\
                                            &    & TOF(a,d,b) CNOT(a,d) NOT(a) NOT(b) NOT(c) TOF4(b,c,d,a) CNOT(b,c) TOF(a,d,c)\\ \hline 
[3,1,7,13,11,0,8,15,2,5,10,6,9,14,12,4]  & 48 & NOT(b) CNOT(b,a) TOF(a,b,c) TOF(a,d,b) CNOT(c,d) TOF4(b,c,d,a) TOF4(a,b,c,d) \\
                                            &    & CNOT(a,c) CNOT(c,b) TOF(b,d,c) NOT(a) NOT(b) CNOT(c,d) CNOT(d,a) TOF(a,b,c)\\ \hline
[3,1,11,7,8,0,9,5,2,6,15,13,14,4,10,12]  & 24 & CNOT(c,b) CNOT(a,d) CNOT(d,a) TOF4(a,b,c,d) TOF(a,b,c) TOF(b,c,a) TOF(a,d,b) \\
                                            &    & CNOT(b,c) NOT(d) NOT(c) NOT(a) TOF(c,d,b) TOF(b,c,d) CNOT(d,c) CNOT(a,c)\\ \hline
[3,5,11,1,8,0,9,7,2,6,14,13,10,4,12,15]  & 24 & CNOT(c,b) TOF(b,d,a) CNOT(a,d) CNOT(d,c) TOF(b,c,a) TOF(a,c,b) TOF(a,d,c) \\
                                            &    & TOF(b,c,a) NOT(d) NOT(c) NOT(b) CNOT(d,a) TOF(b,c,d) CNOT(d,b) TOF(a,b,c)\\ \hline
\end{tabular}
}
\end{table*}

As follows from the previous subsection, there are only five reduced permutations requiring the maximal number of gates, 15. 
We list all five canonical representatives, together with their optimal implementations, in Table \ref{tab:15shki}. Columns of this 
table report the function specification, the number of symmetries this specification generates, and an optimal circuit found by 
our program, correspondingly. The remaining 139 ($=144-5$) permutations requiring 15 gates in an optimal implementation may be found via 
reducing them to a canonical representative through an input/output relabeling and possible inversion.  For example, 
[6,8,15,13,4,0,12,1,3,9,11,14,10,2,5,7] is a permutation requiring 15 gates in an optimal implementation.  It may be obtained from the third listed 
in the Table \ref{tab:15shki} via inversion and relabeling $(a,b,c,d) \mapsto (c,a,b,d)$.

\subsection{Optimal linear circuits}
\begin{table}[h]
\centering
\caption{Number of 4-bit linear reversible functions requiring 0..10 gates 
in an optimal implementation.}
\label{tablin}
\begin{tabular}{|r|r|} \hline
Size  	& Functions  \\ \hline
10		& 138 \\
9		& 13,555 \\
8		& 84,225 \\
7		& 118,424 \\
6		& 72,062 \\
5		& 26,182 \\
4		& 6,589 \\
3		& 1,206 \\
2		& 162 \\
1		& 16 \\
0   	& 1 \\ \hline
\end{tabular}
\end{table}

Linear reversible circuits are the most complex part of quantum error correcting circuits \cite{ar:ag}. Efficiency of these 
circuits defines the efficiency of quantum encoding and decoding error correction operations.
Linear reversible functions are those whose positive polarity Reed-Muller polynomial has only linear terms. More simply, 
and equivalently, linear reversible functions are those computable by circuits with NOT and CNOT gates. 

For example, the reversible mapping 
$a,b,c,d \mapsto b\oplus 1,a \oplus c\oplus 1,d\oplus 1,a$ is a linear reversible function. Interestingly, this linear function is one of the 138 
most complex linear reversible functions---it requires 10 gates in an optimal implementation. The optimal implementation of this function 
is given by the circuit CNOT(b,a) CNOT(c,d) CNOT(d,b) NOT(d) CNOT(a,b) CNOT(d,c) CNOT(b,d) CNOT(d,a) NOT(d) CNOT(c,b). 

We synthesized optimal circuits for all 322,560 4-bit linear reversible functions using FINDOPT algorithm.  
This process took under two seconds on LPTP. The distribution 
of the number of functions requiring a given number of gates is shown in Table \ref{tablin}.

\subsection{Synthesis of Benchmarks}

\begin{table*}[t]
\centering
\caption{Optimal implementations of benchmark functions.}
\label{tab3}
{\footnotesize 
\begin{tabular}{|r|r||r|r|r||r|r|r|} \hline
Name       & Specification         & SBKC  & Source          & PO?   & SOC   & Our optimal circuit & Runtime \\  \hline

4\_49      & [15,1,12,3,5,6,8,7,   & 12    & \cite{www:rlsb} & No    & 12   & NOT(a) CNOT(c,a) CNOT(a,d) TOF(a,b,d)  &  .000355s  \\
           & 0,10,13,9,2,4,14,11]  &       &                 &       &      & CNOT(d,a) TOF(c,d,b) TOF(a,d,c) TOF(b,c,a)  &   \\
             &&&&&                                                          & TOF(a,b,d) NOT(a) CNOT(d,b) CNOT(d,c) & \\\hline

4bit-7-8   & [0,1,2,3,4,5,6,8,7,9, & 7     & \cite{co:m}     & No    & 7    & CNOT(d,b) CNOT(d,a) CNOT(c,d) TOF4(a,b,d,c) & .000002s \\
           & 10,11,12,13,14,15]    &       &                 &       &      & CNOT(c,d) CNOT(d,b) CNOT(d,a) &   \\\hline

decode42   & [1,2,4,8,0,3,5,6,7,9, & 11    & \cite{ar:gaj}   & No    & 10   & CNOT(c,b) CNOT(d,a) CNOT(c,a) TOF(a,d,b)  &  .000004s \\
           & 10,11,12,13,14,15]    &       &                 &       &      & CNOT(b,c) TOF4(a,b,c,d) TOF(b,d,c)  &   \\
             &&&&&                                                          & CNOT(c,a) CNOT(a,b) NOT(a) & \\\hline

hwb4       & [0,2,4,12,8,5,9,11,1,& 11     & \cite{www:rlsb} & Yes   & 11   & CNOT(b,d) CNOT(d,a) CNOT(a,c) CNOT(c,d) & .000052s \\
           & 6,10,13,3,14,7,15]   &        &                 &       &      & TOF(a,d,b) TOF(b,c,a) CNOT(d,c) CNOT(c,b) &   \\
           &&&&&                                                            & TOF(a,c,b) CNOT(a,c) CNOT(b,d) & \\\hline

imark      & [4,5,2,14,0,3,6,10,  & 7      & \cite{ar:pspmh} & No    & 7    & TOF(c,d,a) TOF(a,b,d) CNOT(d,c) CNOT(b,c) & .000003s   \\
           & 11,8,15,1,12,13,7,9] &        &                 &       &      & CNOT(d,a) TOF(a,c,b) NOT(c) &   \\\hline

mperk      & [3,11,2,10,0,7,1,6,  & 9*     & \cite{co:m},    & No    & 9    & NOT(c) CNOT(d,c) TOF(c,d,b) TOF(a,c,d) & .000003s \\
           & 15,8,14,9,13,5,12,4] &        & \cite{ws:p}     &       &      & CNOT(b,a) CNOT(d,a) CNOT(c,a) CNOT(a,b) &   \\
           &&&&&                                                            & CNOT(b,c) & \\\hline

oc5        & [6,0,12,15,7,1,5,2,4,& 15     & \cite{co:szs}   & No    & 11   & TOF(b,d,c) TOF(c,d,b) TOF(a,b,c) NOT(a) & .000158s \\
           & 10,13,3,11,8,14,9]   &        &                 &       &      & CNOT(d,b) CNOT(c,a) CNOT(a,c) TOF(a,b,d) & \\
           &&&&&                                                            & CNOT(c,a) CNOT(c,b) TOF4(a,b,d,c) & \\\hline

oc6        & [9,0,2,15,11,6,7,8,  & 14    & \cite{co:szs}    & No    & 12   & TOF4(a,b,c,d) TOF(b,d,c) CNOT(d,a) TOF(b,c,d) & .000380s \\
           & 14,3,4,13,5,1,12,10] &       &                  &       &      & CNOT(c,b) CNOT(b,c) TOF(a,d,c) TOF(b,c,a) & \\
             &&&&&                                                          & TOF(a,b,c) NOT(a) CNOT(d,b) CNOT(a,d) & \\\hline

oc7        & [6,15,9,5,13,12,3,7,  & 17   & \cite{co:szs}    & No    & 13   & CNOT(b,d) NOT(b) TOF(a,b,c) TOF(b,d,a) TOF(c,d,b) & .0194s \\
             & 2,10,1,11,0,14,4,8] &      &                  &       &      & CNOT(a,d) CNOT(a,c) CNOT(b,a) TOF4(a,b,c,d) & \\
             &&&&&                                                          & TOF(c,d,b) CNOT(c,a) NOT(a) CNOT(b,c) & \\\hline

oc8        & [11,3,9,2,7,13,15,14, & 16   & \cite{co:szs}    & No    & 12   & CNOT(a,b) TOF(b,c,a) TOF(c,d,b) CNOT(d,a) & .000725s \\
             & 8,1,4,10,0,12,6,5]  &      &                  &       &      & TOF4(a,b,d,c) TOF(a,b,d) NOT(b) TOF(a,d,b) & \\
             &&&&&                                                          & TOF(b,d,a) TOF(b,c,d) NOT(a) CNOT(a,d) & \\\hline

nth\_pri   & [0,2,3,5,7,11,13,1,4, & N/A  & N/A              & N/A    & 11  & TOF(a,b,c) CNOT(d,b) TOF(a,c,b) TOF(b,d,c)   & 0.000095s \\
me4\_inc   & 6,8,9,10,12,14,15]    &      &                  &        &     & TOF(b,c,d) CNOT(a,b) TOF4(b,c,d,a) CNOT(c,b) & \\
               &&&&&                                                        & TOF4(a,b,d,c) CNOT(b,a) TOF(b,d,a)           & \\\hline

rd32       & [0,7,6,9,4,11,10,13,  & 4    & \cite{ar:f}      & Yes    & 4   & TOF(a,b,d) CNOT(a,b) TOF(b,c,d) CNOT(b,c) &  .000001s \\
           & 8,15,14,1,12,3,2,5]   &      &                  &        &     &  &   \\ \hline

shift4     & [1,2,3,4,5,6,7,8,9,   & 4    & \cite{co:m}      & Yes    & 4   & TOF4(a,b,c,d) TOF(a,b,c) CNOT(a,b) NOT(a) & .000002s \\
           & 10,11,12,13,14,15,0]  &      &                  &        &     &  &   \\ \hline
\end{tabular}
}
\end{table*}

In this subsection, we report optimal circuits for benchmark functions that have been previously reported in the literature. Table \ref{tab3} summarizes 
the results.  The table describes the {\bf Name} of the benchmark function, its complete {\bf Specification}, {\bf S}ize of the {\bf B}est {\bf K}nown 
{\bf C}ircuit ({\bf SBKC}), the {\bf Source} of this circuit, indicator of whether this circuit has been {\bf P}roved {\bf O}ptimal ({\bf PO?}), {\bf 
S}ize of an {\bf O}ptimal {\bf C}ircuit ({\bf SOC}), the optimal implementation that our program found, and the runtime our program takes to find this 
optimal implementation.  We used the head node of CLSTR for this test, and report the runtime it takes after hash table with all optimal implementations 
with up to 9 gates is loaded into RAM.  Shorter runtimes were identified using multiple runs of the search to achieve sufficient accuracy.  Please note 
that we introduce the function $nth\_prime4\_inc$, which cannot be found in the previous literature. Also, the 9-gate circuit for the function {\em mperk} 
reported in \cite{ws:p} requires some extra SWAP gates to properly map inputs into their respective outputs, indicated by an asterisk.

%

\section{Conclusions and Possible Extensions} \label{s:cfr}

In this paper, we described two algorithms: first, FINDOPT, finds an optimal circuit for any 4-bit reversible function, and second, SEARCALL, finds all 
optimal 4-bit reversible circuits. Our goal was to minimize the number of gates required for function implementation. Our implementation of FINDOPT takes 
approximately 3 hours to calculate all optimal implementations requiring up to 9 gates, and then an average of about $0.00756$ seconds to search for an 
optimal circuit of any 4-bit reversible function. Our implementation of SEARCHALL requires about about 13 days, however, it needs to be completed only 
once to collect all relevant statistics and data. Both calculations are surprisingly fast given the size of the search space.

Using BFS, we demonstrated the synthesis of 117,798,040,190 optimal circuits in 9,688 seconds, amounting to an average speed of 12,168,356 circuits per 
second.  This is over 70 times faster and some 4,500 times more than the best previously reported result (26 million circuits in 152 
seconds)~\cite{ar:pspmh}. Furthermore,  using FINDALL, we demonstrated the synthesis of 20,922,789,888,000 functions in 1,130,276 seconds (18,511,222 
circuits per second). This is over 100 times faster and over 800,000 times more than in \cite{ar:pspmh}. 

We also demonstrated that the search for any given optimal circuit can be done very quickly---$.00756$ seconds per a random function. For example, if all 
optimal circuits were written into a {\em hypothetical} 100+TB 5400 RPM hard drive, the average time to extract a random circuit from the drive would be 
expected to take on the order of $0.01-0.02$ seconds (typical access time for 5400 RPM hard drives).  In other words, it would take longer to read the 
answer from a {\em hypothetical} hard drive than to compute it with our implementation.  Furthermore, the 3-hour calculation of all optimal circuits with 
up to 9 gates could be reduced by storing its result (computed once for the entirety of the described search and its follow up executions) on the hard 
drive, as was done in Subsection \ref{ss:1}. It took 1667 seconds, i.e., under 28 minutes, to load optimal circuits with up to 9 gates into RAM using 
CLSTR. Given that the media transfer rate of modern hard drives is 1Gbit/s (=1GB in 8 seconds) and higher, it may take no longer than 5 minutes ($=300$s 
$> 296=37*8$s) to load optimal implementations into RAM to initiate the search on a different machine.

Minor modifications to the algorithm could be explored to address other optimization issues.  For example, for practicality, one may be interested in 
minimizing depth.  This may be important if a faster circuit is preferred, and/or if quantum noise has a stronger constituent with time, than with the 
disturbance introduced by multiple gate applications.  It may also be important to account for the different implementation costs of the gates used 
(generally, NOT is much simpler than CNOT, which in turn, is simpler than Toffoli). Both modifications are possible, by making minor changes to the first 
part of FINDOPT, and minor modifications of SEARCHALL. To optimize depth, one needs to consider a different family of gates, where, for instance, sequence 
NOT$(a)\;$CNOT$(b,c)$ is counted as a single gate. To account for different gate costs, one needs to search for small circuits via increasing cost by one 
(assuming costs are given as natural numbers), as opposed to adding a gate to all maximal size optimal circuits.

It is also possible to extend the search to find optimal implementations in restricted architectures (see the Section \ref{Extension} for details). 
Finally, the search could be extended to find some small optimal 5-bit circuits.  A simple calculation shows that $80^6\log_{2}{(80)}/5!/2$ bits (the 
number of elementary transformations to the power of depth, times space to store a single gate, divided by the number of symmetries) suffices to store all 
optimal circuits containing up to 6 gates for 5-bit permutations. Thus, a search of optimal implementations may be carried to compute optimal circuits 
with up to 12 gates. However, it is possible that a larger search may be performed.

Finally, techniques reported in this paper may be applied to the synthesis of optimal stabilizer circuits. Coupled with peep-hole optimization algorithm 
for circuit simplification, these results may become a very useful tool in optimizing error correction circuits. This may be of a particular practical 
interest since implementations of quantum algorithms may be expected to be dominated by the error correction circuits. 

\section{Extension: LNN circuits} \label{Extension}
Of all possible extensions of the presented search algorithms described in Section \ref{s:cfr}, the most 
computationally difficult is the one where the underlying architecture is restricted.  This is because 
the number of input/output labeling symmetries that can be used to reduce the search is equal to the number 
automorphisms of the unlabeled graph corresponding to the underlying architecture.  For the complete graph on four
bits, $K_4$, its number of automorphisms is maximal, $4!=24$.  Since the number of automorphisms is maximized, 
this has helped us to gain maximal advantage.  Of the connected graphs with four nodes, the chain, 
corresponding to the LNN (Linear Nearest Neighbor) architecture, 
has the least number of automorphisms, being just two.  In this section we will illustrate 
that our search may be modified to find optimal circuits in the LNN architecture, implying that it is at least as 
efficient for the remaining four possible architectures (the number of non-isomorphic unlabeled connected graphs 
on four nodes minus two, one for the LNN and one for $K_4$). 

The restriction to the LNN architecture implies that the gate set is limited to those gates operating on the subset of 
qubits that is a continuous substring of the string of all variables, $abcd$.  For example, gates CNOT$(a,b)$ and 
TOF$(b,d,c)$ are allowed, and gate TOF$(a,b,d)$ is disallowed.  Such restriction to the LNN architecture does not 
necessarily imply direct physical applications.  In fact, not only it is not certain that the underlying architecture 
is LNN (and, it must be noted that local architecture may differ from global architecture), and not only do we not account 
for the individual gate costs, but the restriction itself may not be physically grounded.  Indeed, according to 
\cite{ar:mm}, LNN-optimal NCV implementation of the TOF$(a,c,b)$ requires 13 gates and depth 13; however, TOF$(a,b,c)$ requires 
only 9 gates and depth 9, and TOF$(a,b,d)$ may be implemented with 15 NCV gates and depth of only 12.  This means that TOF$(a,b,d)$ 
is ``faster'' than TOF$(a,c,b)$, and of the two we have just disallowed TOF$(a,b,d)$.  In other words, we suggest that our 
software is updated to achieve the results relevant to experiments once all physical restrictions are known; however, 
the goal of this paper, and this Section, in particular, is to illustrate that when needed proper modifications are possible.

We have modified implementation of the FINDOPT algorithm to account for restrictions imposed by the LNN architecture.  This required 
to change the definition of the equivalence class of a 4-bit reversible function.  In particular, the newly defined LNN-equivalence class 
allows symmetries with respect to the inversion, and one of the two possible relabelings: $(a,b,c,d) \mapsto (a,b,c,d)$ and 
$(a,b,c,d) \mapsto (d,c,b,a)$.  The remainder of code and algorithms remained essentially the same. 

We report the result of the optimal LNN synthesis in the following three tables, Table \ref{tab2lnn}, Table \ref{tab1lnn} and Table \ref{tab3lnn}.  The 
data contained is analogous to that reported in Table \ref{tab2}, Table \ref{tab1}, and Table \ref{tab3} correspondingly.  It took the total of 
1,313,020.23 seconds to calculate 10,000,000 LNN-optimal reversible circuits reported in the Table \ref{tab2lnn}.  This calculation has been performed on 
three cluster nodes in parallel, reducing physical time spent on this calculation by a factor of three.  The average time to calculate a single random 
LNN-optimal 4-bit reversible circuit is approximately 0.131 seconds.  Based on the number and distribution of circuits in Tables \ref{tab1} and 
\ref{tab1lnn} we conjecture that there are no LNN-optimal circuits requiring 21 gates, and as such it suffices to have generated all optimal circuits with 
up to 10 gates to synthesize any LNN-optimal 4-bit reversible circuit. 

\begin{table}[t]
\centering
\caption{Distribution of the number of gates required for LNN-optimal implementation of 10,000,000 random 4-bit reversible functions.}
\label{tab2lnn}
\begin{tabular}{|r|r|} \hline
Size  	& Functions 		\\  \hline
18		& 6 	\\
17		& 20,546 	\\
16		& 1,091,953 	\\
15		& 3,976,746 	\\
14		& 3,286,497 	\\
13		& 1,244,670 	\\
12		& 308,993		\\
11		& 59,289 		\\
10		& 9,693 		\\
9		& 1,387			\\
8		& 189 			\\
7		& 26 				\\
6		& 5				\\ \hline
\end{tabular}
\end{table}

\begin{table}[t]
\centering
\caption{Number of 4-bit permutations requiring prescribed number of gates in the LNN architecture.}
\label{tab1lnn}
\begin{tabular}{|r|r|r|r|} \hline
Size  	& Functions 		 & Reduced         &  Runtime \\
      	&           		 & Functions       &          \\ \hline
$\geq$11& Unknown			 & Unknown         &            \\
10		& 20,355,134,386	 & 5,089,090,158   & 13,299.9s \\
9		& 2,921,376,642 	 & 730,451,187	   & 1,642.72s \\
8		& 378,041,753 	 	 & 94,551,844 	   & 241.367s \\ 
7		& 44,754,539 		 & 11,201,218 	   & 68.192s    \\ 
6		& 4,886,991 		 & 1,226,080 	   & 11.041s    \\
5		& 493,788 			 & 124,628 		   & 1.28s     \\ 
4		& 46,108 			 & 11,885 		   & 0.18s      \\ 
3		& 3,947 			 & 1,083 		   & 0.02s     \\
2		& 303 				 & 100			   & $<$0.001s  \\
1		& 20				 & 10			   & $<$0.001s  \\
0   	& 1     			 & 1			   & $<$0.001s  \\ \hline
Total   & 23,704,738,478     & 5,926,658,194   & 15,264.692s  \\ \hline
\end{tabular}
\end{table}

\begin{table*}[t]
\centering
\caption{LNN-optimal implementations of benchmark functions.}
\label{tab3lnn}
{\footnotesize 
\begin{tabular}{|r|r|r|r|} \hline
Name       	& \# gates & LNN optimal circuit & Runtime \\  \hline
4\_49      	& 16 &  CNOT(d,c) TOF(c,d,b) CNOT(b,a) NOT(c) TOF(b,c,d) CNOT(c,b) TOF4(a,b,c,d) TOF4(a,c,d,b) & 0.68s \\
&& 					TOF(a,b,c) CNOT(a,b) NOT(a) CNOT(c,d) TOF(b,d,c) TOF(b,c,a) TOF(a,c,b) NOT(c) &   \\\hline
4bit-7-8   	& 7  &  CNOT(d,c) TOF(c,d,b) TOF4(b,c,d,a) TOF4(a,b,c,d) TOF4(b,c,d,a) TOF(c,d,b) CNOT(d,c) &  $<$0.001s \\\hline
decode42   	& 13 &  NOT(d) NOT(c) CNOT(c,b) TOF(b,d,c) TOF(b,c,a) TOF4(a,c,d,b) TOF4(a,b,c,d) CNOT(d,c) & $<$0.001s \\
&&					TOF(b,c,a) TOF4(a,c,d,b) NOT(d) TOF(c,d,b) NOT(c) &   \\\hline
hwb4       	& 16 &  TOF(a,b,c) CNOT(c,d) CNOT(b,c) TOF(b,c,a) TOF4(a,c,d,b) NOT(a) CNOT(a,b) CNOT(d,c) & 1.34s \\
&& 					CNOT(b,a) TOF(b,c,d) TOF(b,c,a) TOF4(a,c,d,b) TOF(a,b,c) NOT(b) CNOT(c,d) TOF(a,b,c) &   \\\hline
imark      	& 11 &  TOF4(b,c,d,a) TOF4(a,b,c,d) CNOT(b,c) TOF4(a,b,c,d) TOF4(b,c,d,a) CNOT(d,c) TOF(a,c,b) & $<$0.001s \\
&& 					CNOT(b,a) TOF(c,d,b) NOT(c) CNOT(b,a) &   \\\hline
mperk      	& 11 &  CNOT(d,c) TOF(a,c,b) CNOT(b,a) TOF(c,d,b) CNOT(a,b) NOT(c) TOF(b,c,d) CNOT(c,b) & $<$0.001s \\ 
&&					CNOT(b,a) TOF(c,d,b) CNOT(b,c) &   \\\hline
oc5        	& 14 &  CNOT(d,c) TOF4(a,c,d,b) CNOT(c,b) TOF(b,d,c) CNOT(b,a) CNOT(b,c) TOF(b,c,d) CNOT(c,b) & 0.01s \\
&&					NOT(a) CNOT(a,b) CNOT(b,c) TOF(b,c,a) TOF4(b,c,d,a) TOF4(a,b,d,c) &   \\\hline
oc6        	& 14 &  TOF(b,c,a) TOF4(a,c,d,b) TOF4(a,b,c,d) NOT(b) CNOT(b,a) CNOT(d,c) TOF4(b,c,d,a) & 0.01s \\
&& 					CNOT(c,b) TOF(a,b,c) CNOT(c,d) TOF(b,d,c) NOT(b) TOF(a,b,c) TOF(b,c,d) &   \\\hline
oc7        	& 15 &  TOF(b,c,d) TOF(c,d,b) TOF4(b,c,d,a) TOF(a,b,c) NOT(b) TOF4(a,b,d,c) TOF(b,c,d) CNOT(c,b) & 0.07s \\
&&					CNOT(b,a) NOT(d) TOF(c,d,b) TOF4(a,b,d,c) NOT(a) CNOT(c,d) TOF(a,b,c) &   \\\hline
oc8        	& 14 &  CNOT(a,b) TOF(b,d,c) CNOT(c,d) NOT(c) TOF(b,c,a) TOF4(a,b,c,d) CNOT(a,b) CNOT(d,c) & 0.03s \\
&&					CNOT(b,c) TOF4(a,c,d,b) CNOT(c,b) CNOT(b,a) TOF4(a,b,c,d) TOF(b,d,c) &   \\\hline
nth\_prime4	& 11 &  CNOT(d,c) TOF(b,c,a) CNOT(b,c) NOT(b) TOF(b,c,d) TOF(b,c,a) TOF4(a,b,d,c) & $<$0.001s \\
&&					TOF(a,c,b) NOT(a) TOF4(a,c,d,b) CNOT(b,a) &  \\\hline
rd32       	& 7  &  TOF(b,c,d) NOT(c) TOF4(a,b,c,d) NOT(c) CNOT(a,b) TOF4(a,b,c,d) CNOT(b,c) & $<$0.001s  \\\hline
shift4     	& 4  &  TOF4(a,b,c,d) TOF(a,b,c) CNOT(a,b) NOT(a) & $<$0.001s  \\\hline
\end{tabular}
}
\end{table*}

\section{Acknowledgments}

We wish to thank Dr. Donny Cheung from Tornado Medical Systems and Dr. Sean Falconer from Stanford University for their useful discussions and 
contributions.  We wish to thank Marek Szyprowski and Prof. Pawel Kerntopf from Warsaw University of Technology for their comments.

This article was based on work partially supported by the National Science Foundation, during
D.~Maslov's assignment at the Foundation.

\bibliographystyle{abbrv}

\end{document}